\documentclass[prl,twocolumn,superscriptaddress]{revtex4-1}
\pdfoutput=1 
\usepackage{graphicx}
\usepackage{amssymb}
\usepackage{amsmath}
\usepackage{color}
\usepackage{hyperref}
\usepackage{bm}
\usepackage{color}

\begin{document}

\title{Accelerated kinetic over-relaxation for dynamic 
energy minimization in lattice many-body problems}

\author{Nasrollah Moradi}
\thanks{nasrollah.moradi@rub.de}
\affiliation{Freiburg Institute for Advanced Studies (FRIAS), University of Freiburg, Albertstrasse 19, 79104, Freiburg, Germany}
\author{Andreas Greiner}
\affiliation{Department of Microsystems Engineering (IMTEK), University of Freiburg, Georges-Köhler-Allee 103, 79110 Freiburg, Germany}
\author{Simone Melchionna}
\affiliation{PCF-CNR, Piazzale A. Moro, 5, 00185, Roma, Italy }
\author{Francesco Rao}
\affiliation{Freiburg Institute for Advanced Studies (FRIAS), University of Freiburg, Albertstrasse 19, 79104, Freiburg, Germany}
\author{Sauro Succi}
\affiliation{IAC-CNR, via dei Taurini 9, 00185, Roma, Italy}
\affiliation{Freiburg Institute for Advanced Studies (FRIAS), University of Freiburg, Albertstrasse 19, 79104, Freiburg, Germany}

\date{\today}

\begin{abstract}
A kinetic over-relaxation minimizer, based on a  lattice version of the Boltzmann equation is presented.
The method is validated against standard Metropolis Monte Carlo, and proves very effective 
in attaining (global) minima of classical pair potentials, involving solid body rotations.
Linear scaling of the computational time to minimum with the system size is demonstrated. 
\end{abstract}

\maketitle
\paragraph{   Introduction.---}  Finding the minimum energy configuration in classical and quantum many-body
systems is crucial for many applications in science and engineering \cite{Kirkpatrick1983,Marinari1992,Bryngelson11995}.
The highly-dimensional configuration-space of many-body systems confines analytical results
to very special cases \cite{Lieb1968}, and also commands
major extensions of  standard minimization tools from linear and non-linear algebra. 
To this regard, dynamic minimization procedures in real or
fictitious time, such as Monte Carlo (MC) \cite{MCbook1,MCbook2}  or Langevin Molecular Dynamics (LMD) \cite{Nosea1984,Car1985}
and their modern variants, have proven very effective for the last decades.
Here, we present a new kind of dynamic minimizer, which is based on a 
lattice version of Boltzmann's kinetic equation, Lattice Boltzmann (LB) for short. 
The method is applicable to a class of lattice potentials involving solid body rotations.
The LB propagation-relaxation dynamics proves
very effective in reaching the (global) minimum energy configuration of lattice many-body 
potentials.  The time to minimum is found to be a fraction of a standard
Metropolis MC and, perhaps more importantly, it is found to scale linearly with the system size.

\begin{figure*} [t]
(a) \includegraphics[width=44mm]{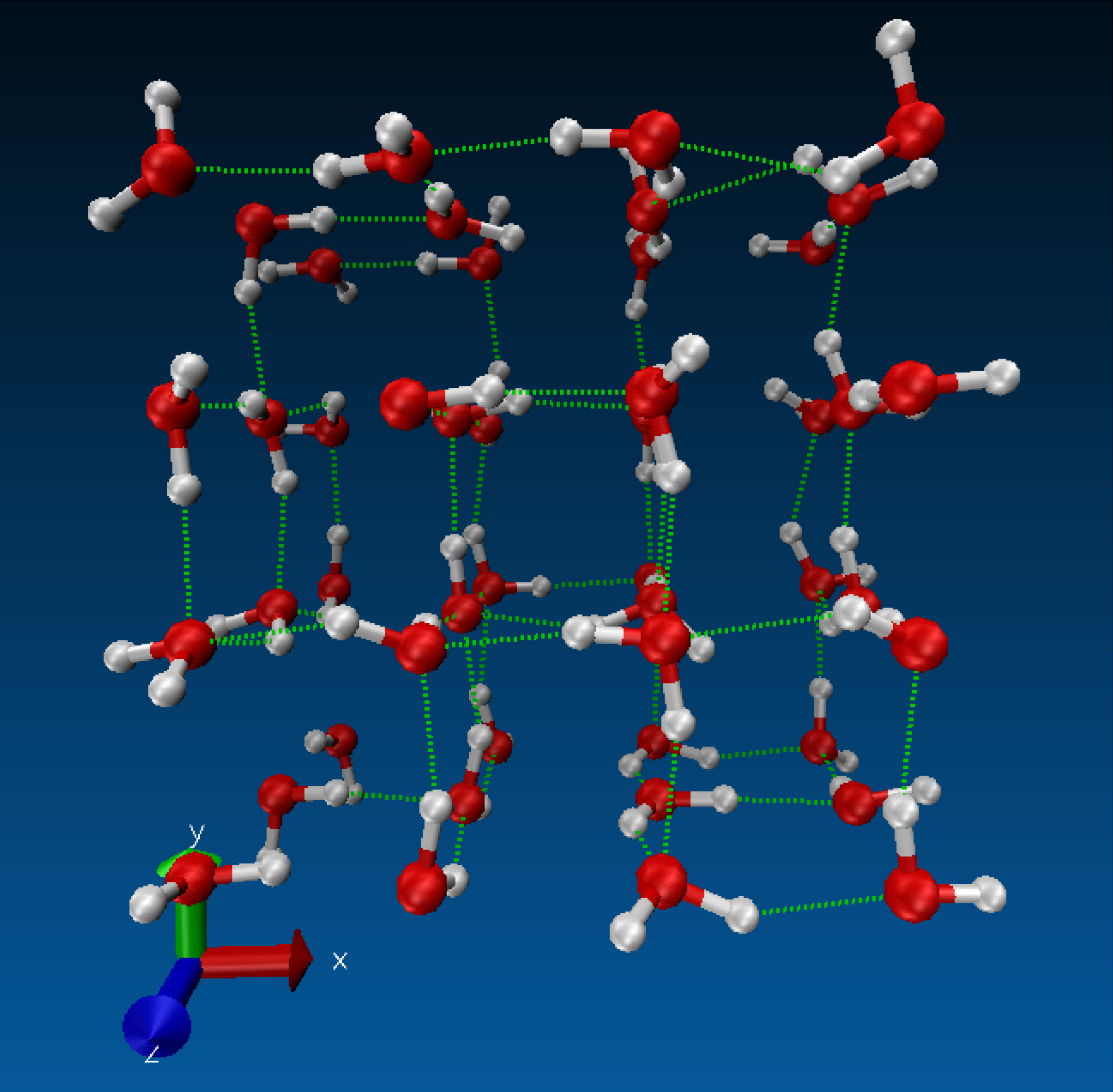}\hspace{10mm}
(b)\includegraphics[width=44mm]{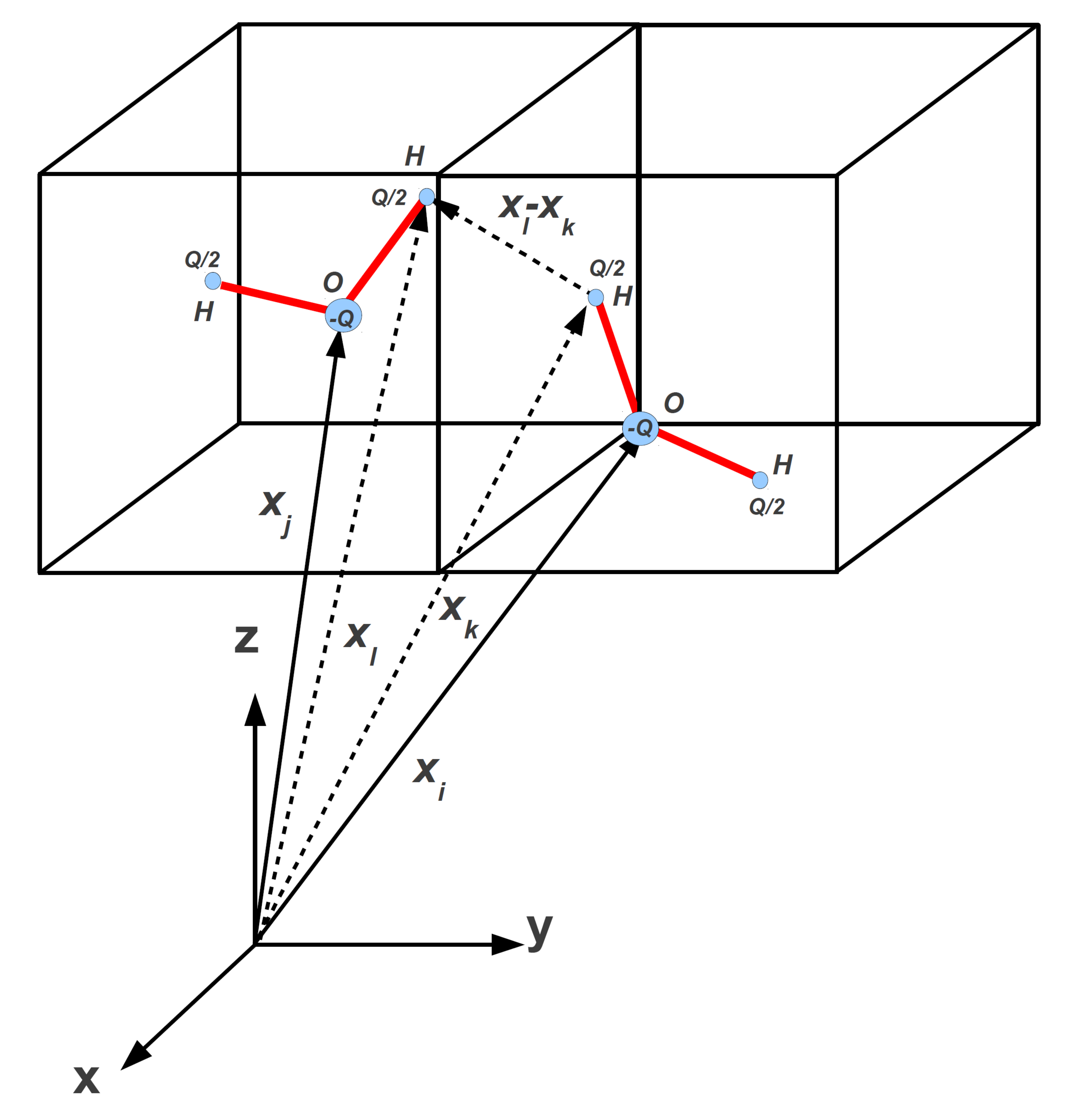} 
(c)\includegraphics[width=44mm]{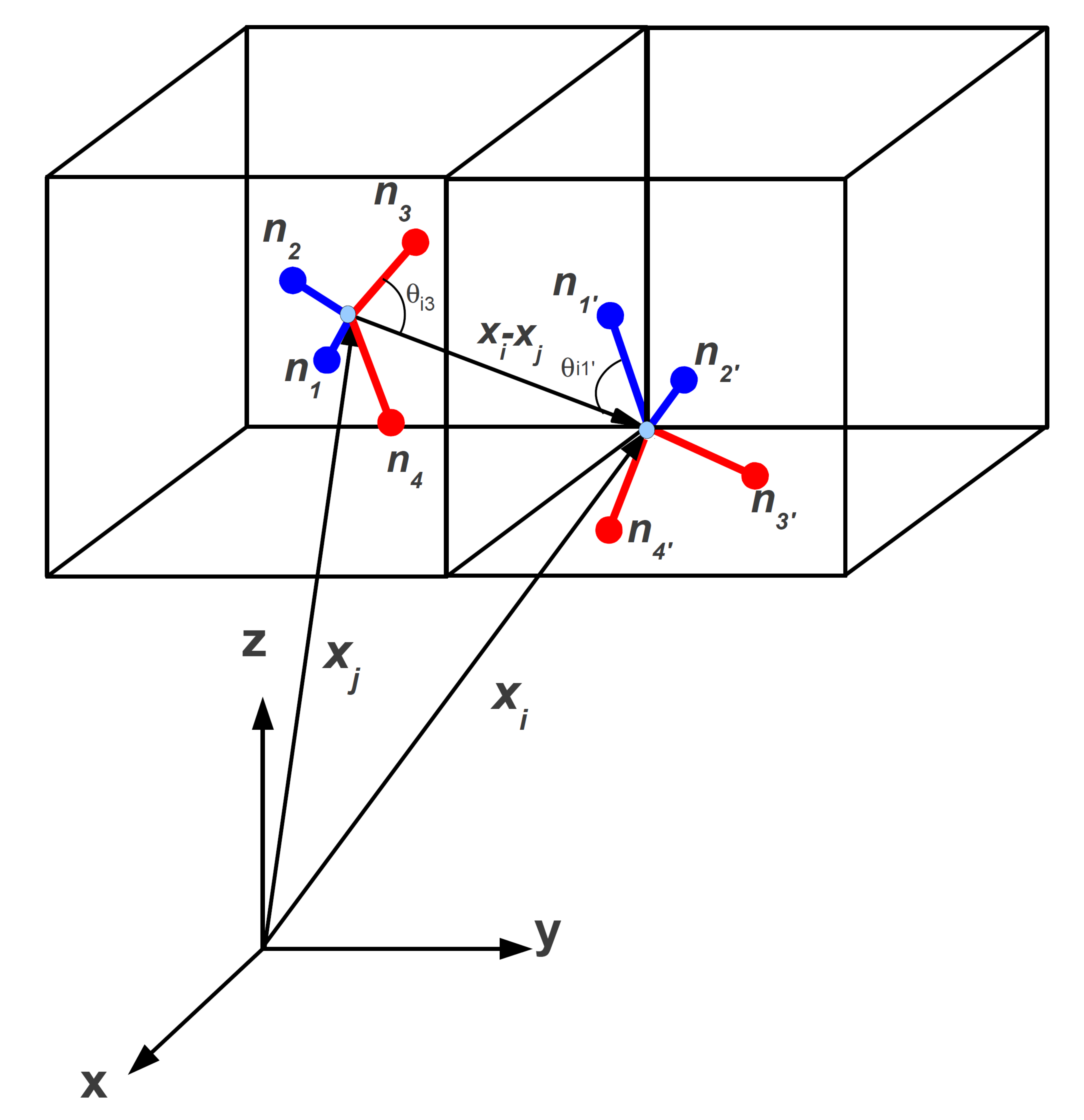}
\caption{ Simulation models. (a): A typical configuration of $\text{H}_{2}\text{O}$ molecules in the lattice TIP3P  model. 
The dashed lines show the hydrogen bonds. Here, for the sake of visibility, a small system, $N=4\times4\times3$, is presented. (b): Geometry of two neighbor $\text{H}_{2}\text{O}$ molecules in lattice TIP3P  model. The  molecules are considered as a solid object with two arms with the angle of $104.52^\circ$ (c): Tetrahedral geometry of two neighbor $\text{H}_{2}\text{O}$ molecules in lattice BN3d model, a solid object with  four arms representing two donors and two acceptors  denoted by the corresponding normals $n_k$, $k=1,4$ (unprimed) and $n_l$, $l=1,4$ (primed),  respectively.}
\label{fig:H2O}
\end{figure*}
\paragraph{ The physical problem.---}  We represent water molecules by solid bodies, sitting on the nodes of a regular lattice, consisting of a finite number of
point particles carrying either mass and/or charge, and moving along the links of the lattice at discrete times. 
Here, we consider the standard so-called D3Q27 lattice, consisting of the $26$ neighbors of the center node
of a cubic cell (6 face centers, 12 edge centers and 8 vertices). 
In fig.~\ref{fig:H2O} (a) a typical configuration of   water  molecules on the lattice is shown.  The centers of mass (COM) of the water molecules are fixed at the lattice nodes. The molecules are free to rotate about their COM under the action of a torque arising from the intermolecular interaction.  Each molecule interacts with its 
26 neighbors and its orientation is characterized by the quaternion technique \cite{Rapaport:1985kc}. In this way, in order to determine the rotation matrix one needs 
the time derivative of the components of the  quaternion which are functions of the angular velocity $\mathbf{\Omega}$.  
The angular velocity evolves in time under the effect of the torque: 
\begin{equation}
\mathbf{T}(\mathbf{x})=-\partial_{q(\mathbf{x})}\sum_{\mathbf{y}}V\left(q(\mathbf{x}),q(\mathbf{y})\right)
\label{TORQUE}
\end{equation}
where the angular potential $V\left(q(\mathbf{x}),q(\mathbf{y})\right)$ is a function of the
orientations $q(\mathbf{x})$ and $q(\mathbf{y})$ of the interacting molecules located at $\mathbf{x}$ and $\mathbf{y}$. 
In this paper, we shall explore two specific lattice potentials for the water based on the  TIP3P   \cite{Jorgensen,pi2009anomalies}  and 
the three-dimensional Ben-Naim (BN3d)  \cite{BNbook} potential, their respective geometry being given in fig.~\ref{fig:H2O}~(b) and (c), respectively. 

\begin{figure}
\includegraphics[width=42mm]{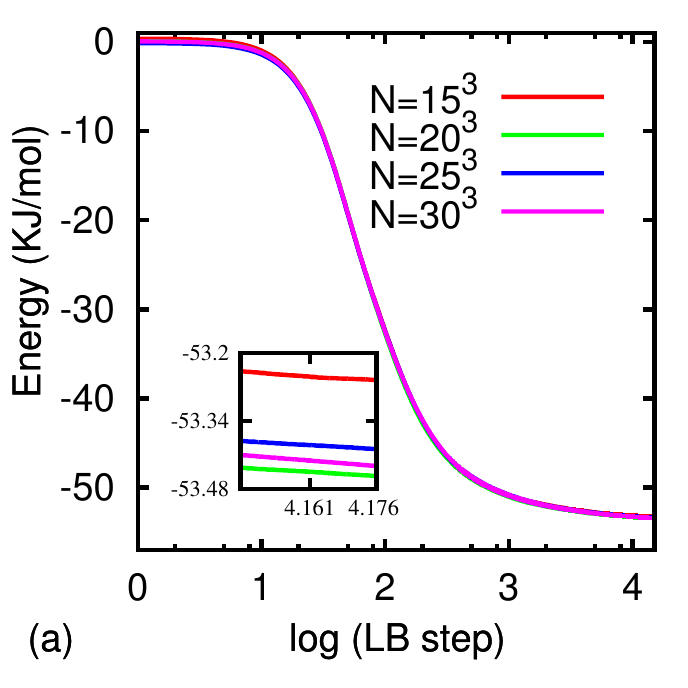}
\includegraphics[width=42mm]{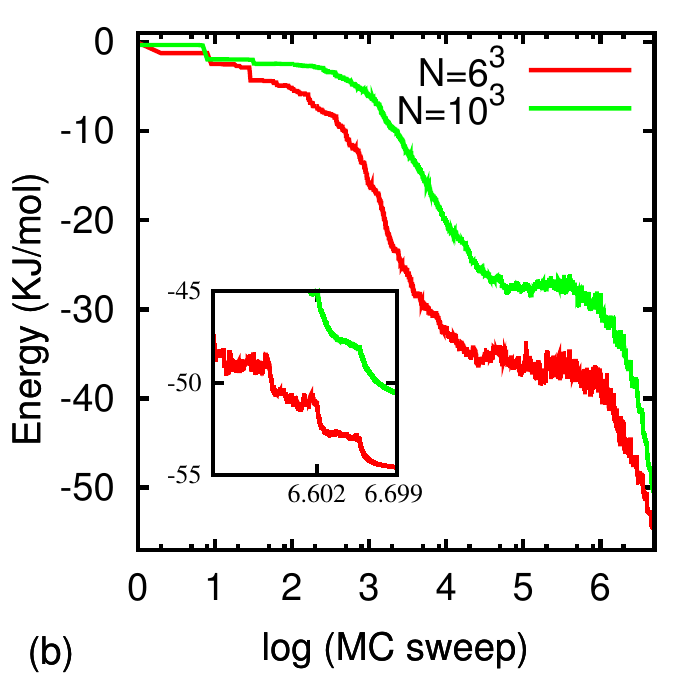}
\caption{(a): Potential energy for the water model based on the LTIP3P versus LB steps with $\tau/\Delta t=20$ (over-relaxation regime) and $\gamma=50 \Delta t^{-1}$ for different system sizes. (b):  Potential energy versus (simulated annealing) MC sweeps for two different system sizes. The insets zoom into final energies.}
\label{fig:MCLB}
\end{figure}

\paragraph{  Lattice TIP3P (LTIP3P) model.---}
 In the LTIP3P model the potential of two neighboring molecules located at lattice nodes $i$ and $j$   is given by   
\begin{equation}
V(\mathbf{x}_i,\mathbf{x}_j) =\frac{1}{4\pi\epsilon}\sum_{k=1}^3\sum_{l=1}^3 \frac{Q_k Q_l}{\lvert\mathbf{x}_l-\mathbf{x}_k\rvert}
\label{eq:LTIP3P}
\end{equation}
in which $\epsilon$ is  the relative permittivity. The indices $k$ and $l$ refer to the charges of the water molecules  associated  with two hydrogens and one oxygen atoms  for each molecule.   Both  hydrogens  carry the  same positive charge  ($+Q/2$) while the  oxygen carries a negative charge ($-Q$). The Lennard-Jones potential is neglected  since   the COM of the water molecules are fixed at the lattice nodes. We use the standard 
parameters for  TIP3P water model given in \cite{Jorgensen}. Besides, the distance between the COM of a molecule with its nearest neighbor  is fixed to  $2.77$ $\text{A}^\circ$, as observed from regular molecular dynamics simulation of the bulk liquid \cite{Mark}. In order to determine the full potential of a single molecule we sum over the first shell of neighboring molecules only (26 molecules).
\paragraph{  Lattice BN3d (LBN3d) model.---} In the LBN3d model  the potential of two neighboring water molecules located at lattice nodes $i$ and $j$  (with the tetrahedral structure carrying two donors and two acceptor)  is given by the BN3d potential  \cite{BNbook} as
\begin{widetext}
\begin{equation}
V(\mathbf{x}_i,\mathbf{x}_j) =  W(\rho) e^{-\frac{1}{2}(\frac{\lvert\mathbf{x}_i-\mathbf{x}_j\rvert-R_\text{HB}}{{\sigma_R})^2}}\sum_{k=1}^4\sum_{l=1}^4 \epsilon_{kl}^{HB}  e^{-\frac{\left(\hat{\mathbf{n}}_k \cdot \hat{\mathbf{r}}_{ij}-1\right)^2+\left(\hat{\mathbf{n}}_l \cdot \hat{\mathbf{r}}_{ij}+1\right)^2}{2\sigma_{\theta}^2}}.
\label{eq:LBN3dP}
\end{equation}
\end{widetext}
In Eg.~(\ref{eq:LBN3dP})   $\epsilon_{kl}^{HB}=\mp1$ is a selective matrix, the plus sign for the repulsive (donor-donor and acceptor-acceptor) and the minus  for the attractive (donor-acceptor) interactions. The density factor $W(\rho)$ is a coupling strength  set to $1$, $R_\text{HB}$ is the selected length of the hydrogen bond, $\sigma_R$ controls the sharpness of the radial interaction and $\sigma_\theta$ is a parameter adjusting the the stiffness of directional interactions \cite{Mazzitelli:2011uv}.  The unit vector  in the direction of the tetrahedral arm $k$  is denoted by  $\hat{\mathbf{n}_k}$ and  $\hat{\mathbf{r}}_{ij}=(\mathbf{x}_i-\mathbf{x}_j)/|\mathbf{x}_i-\mathbf{x}_j|$ is the unit vector along the  link of the two lattice nodes  $i$ and $j$. Here,  as in the case of LTIP3P, each  molecule interacts with its all neighbors in the first shell.
\paragraph{ The kinetic equation for the quaternion moments.---}
The equation of motion of the orientational degrees of freedom is most
conveniently cast in quaternion  form \cite{Rapaport:1985kc}, namely 
\begin{equation}
\label{Q}
\partial_t q_{\mu} + \mathbf{u} \cdot \nabla q_{\mu} = D \Delta q_{\mu} + \dot q_{\mu}
\end{equation}
where $\mu=0,3$ is the quaternion index, $\mathbf{u}$ is the fluid velocity, $D$
the translational diffusivity and $\dot q_{\mu}$ the drive due to the torque \cite{Moradi2013}.  The above equation describes the
transport of the angular momentum across the fluid, plus the drive due to the
local torque.  The local angular momentum  $\mathbf{L}=\mathbf{I} \cdotp\mathbf{\Omega}$ obeys a
Langevin equation of the form
\begin{equation}
\label{LANGE}
\mathbf{I} \cdotp \frac{d \mathbf{\Omega}}{dt} = -  \gamma \mathbf{I} \cdotp \mathbf{\Omega} + \mathbf{T} 
\end{equation}
where $\mathbf{I}$ is the inertia tensor of the molecule, $\gamma$ is an
effective friction, while the torque is given in Eq.~(\ref{TORQUE}).  Eqs.~(\ref{Q}) and
(\ref{LANGE}) are evolved until the (global) minimum energy configuration is
reached.  To this purpose, the total energy of the system $V(t)=\sum_{\mathbf{x},\mathbf{y}}
V(q(\mathbf{x};t),q(\mathbf{y};t))$ is monitored in time.  It has been shown that the
decay of the total energy is associated to the dynamic formation of hydrogen
bonds between the water-like molecules \cite{BNbook,Moradi2013}.  
Hence, the number of hydrogen bonds per molecule (HBs) serves as a 
representative order parameter to describe the energy
landscape of this classical many-body system. 
\paragraph{ The computational method.---}
The Langevin equation is solved in the over-damped approximation,
to deliver  $\mathbf{\Omega} =  (\mathbf{I}^{-1}\cdot\mathbf{T})/\gamma$. 
The quaternion equation is solved by means of a Lattice Boltzmann 
technique \cite{Succi:2001a, BSV}. 
In this formulation, each lattice node hosts a set of discrete
distributions $q_{i,\mu}$, $i=0,n$, connecting each lattice site to its $n$
neighbors (here $n=26$). The lattice is chosen with sufficient symmetry (D3Q27) to support
macroscopic advection and diffusion phenomena \cite{Moradi2013}.
The LB dynamics reads as follows
\begin{eqnarray}
\label{QLB}
q_{i,\mu} (\mathbf{x}+\mathbf{c}_i \Delta t,t+\Delta t)&=&
(1-\omega) q_{i,\mu}(\mathbf{x},t)  + \omega q_{i,\mu}^{eq}(\mathbf{x},t) \nonumber \\ &+& \dot q_{i,\mu}(\mathbf{x},t)\Delta t
\end{eqnarray}
where $\omega=\Delta t/\tau$ is a dimensionless relaxation parameter, $\tau$
being the typical relaxation time due to ``particle" collisions and $\mathbf{c}_i$ is the particle velocity in the direction of the i-th lattice link.  Such collisions drive the distribution to the following local equilibrium
\begin{equation}
q_{i,\mu}^{eq} = w_i q_{\mu}  (1 +  u_i)
\end{equation}
where $q_{\mu} = \sum_i q_{i,\mu}$ is the quaternion ``density",
$w_i$ are standard lattice weights normalized to unity and $u_i = \mathbf{u} \cdot \mathbf{c}_i/c_s^2$, is the
projection of the local flow velocity upon the i-th lattice link, $c_s\equiv c / \sqrt{3}$ being the
lattice speed of sound given by $c_s^2  = \sum_i w_i c_i^2$.  It should be noted that since we focus on the bulk water (no hydrodynamics) we set $\mathbf{u}$ to zero. To enhance numerical stability we use the so-called ``fractional time stepping"  \cite{Qian1997,Zhang2001}. The Eq.~(\ref{QLB}) then modifies into 
\begin{widetext}
\begin{eqnarray}
\label{FQLB}
q_{i,\mu} (\mathbf{x}+\mathbf{c}_i \Delta t,t+\Delta t)&=& p[(1-\omega) q_{i,\mu}(\mathbf{x},t) +  \omega q_{i,\mu}^{eq}(\mathbf{x},t) + \dot q_{i,\mu}(\mathbf{x},t) \Delta t] 
+(1-p)[(1-\omega) q_{i,\mu}(\mathbf{x}+\mathbf{c}_i\Delta t,t) \nonumber \\ &+& \omega q_{i,\mu}^{eq}(\mathbf{x}+\mathbf{c}_i\Delta t,t) + \dot q_{i,\mu} (\mathbf{x}+\mathbf{c}_i\Delta t,t)\Delta t],
\end{eqnarray}
\end{widetext}
where $p$ ($0<p<1$)  is a parameter fixing the fraction of  the distance traveled
in a single sub-step. 
In our simulations, $p=1/3$ proved adequate to avoid numerical  instabilities.  
\paragraph{ The LB method as a kinetic over-relaxation scheme.---}
It is well known that relaxation methods can be cast in the form of a diffusive
process in fictitious time. Numerical stability of the discretized version imposes
the following  diffusive Courant-Friedrichs-Lewy (CFL) constraint \cite{Courant} on the
relaxation parameter $C_d$
\begin{equation}
\label{CFLD}
C_d \equiv \frac{D \Delta t}{\Delta x^2} < 1/d
\end{equation} 
in $d$ spatial dimensions.
The LB scheme is based on a streaming-relaxation dynamics
in the extended kinetic space, which features some advantageous properties 
from the computational point of view.
First, since the LB dynamics is hyperbolic (first order in both space and time), the stability 
constraint comes in advection rather than diffusion form, namely $ C_a \equiv c \Delta t/\Delta x  \le 1 $.
In fact, LB operates strictly at $C_a=1$, which means that the time-step
scales linearly  with the space resolution $\Delta x$, instead of quadratically
as in a diffusion process \cite{Succi:2001a}. This is a significant advantage as the size of the system is increased.
The stability condition for the relaxation step reads as $ \Delta t < 2 \tau $. This is independent of the lattice spacing, because relaxation is fully local.
The above condition is a statement of positivity of the LB mean free path $ \lambda = c_s (\tau - \Delta t/2)$. Note that while the LB dynamics recovers diffusive behavior in the limit
of low Knudsen number \cite{Succi:2001a}, $\lambda \partial_x q/q \ll 1$  ($q$ is a generic quantity, say any component of the
quaternion),  it is by
no means restricted to this regime. 
In the opposite limit, $Kn \to \infty$, the LB dynamics describes
the free-ballistic regime, which is still a perfectly meaningful 
dynamics for the purpose of minimization.
Remarkably, it is precisely in this quasi-ballistic regime, $Kn \gg 1$, that
the LB dynamics provides its best as a dynamic many-body minimizer.
This stands in sharp contrast with the standard use of LB as a 
hydrodynamic solver in kinetic disguise.
As detailed shortly, in the case of LTIP3P, LB shows optimal performance 
around  $\tau/\Delta t \sim 20$, which, in lattice units, corresponds
to $C_d \sim 20$, an order of magnitude above the CFL limit! 
\begin{figure} [t] 
\includegraphics[width=42mm] {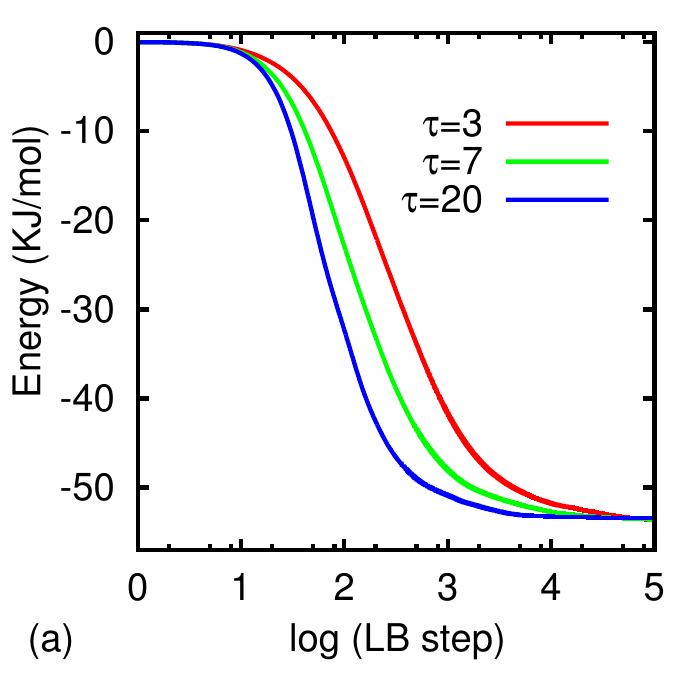}
\includegraphics[width=42mm]{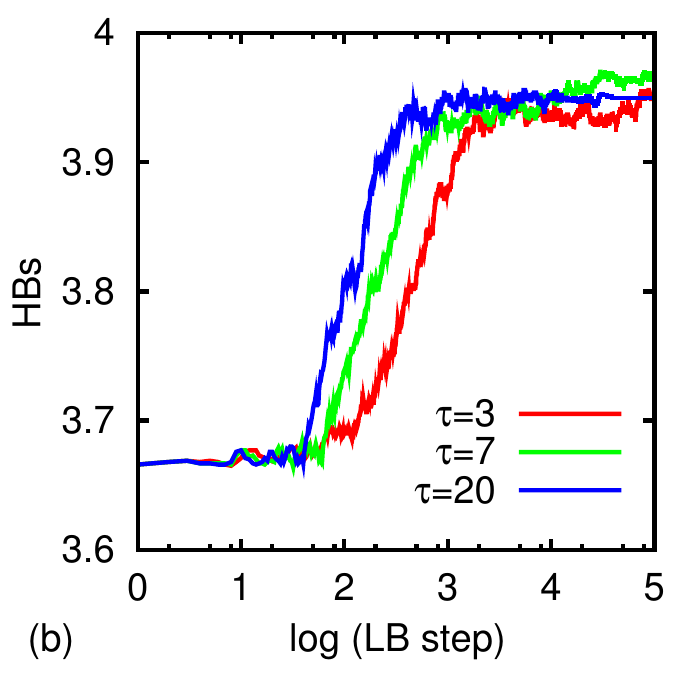}\\
\includegraphics[width=42mm]{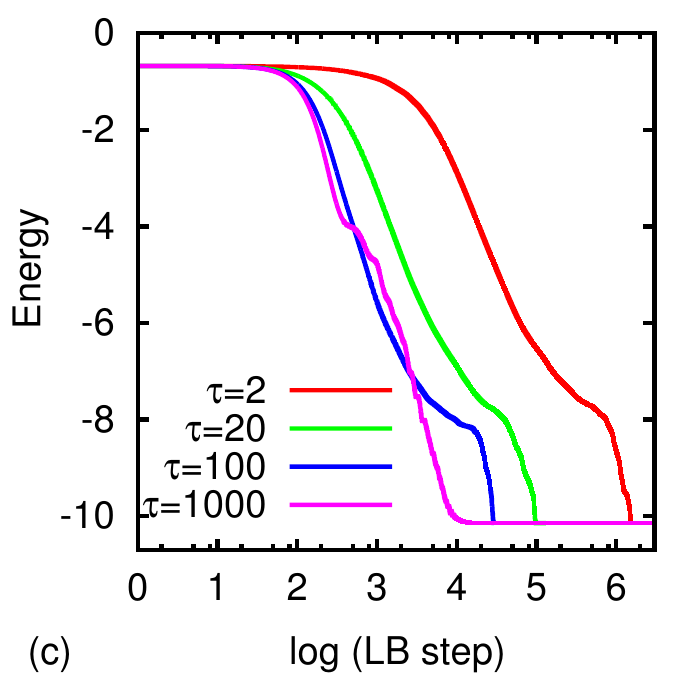} 
\includegraphics[width=42mm]{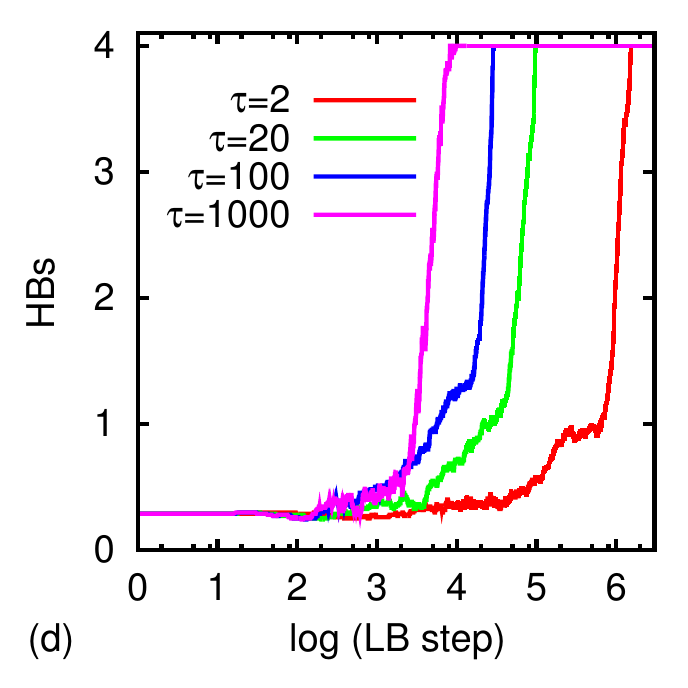}
\caption{Energy vs LB timesteps for the case of  LTIP3P
at different $\tau$ in panel (a) and the corresponding  HBs in panel (b),  with $N=10^3$ and $\gamma=50 \Delta t^{-1}$.
The figures in the lower row report the same information as in the upper row, for the case of LBN3d  with $N=6^3$ and $\gamma=10^5 \Delta t^{-1}$.}
\label{fig:taus}
\end{figure}
\paragraph{ Numerical Results.---}
We have run several simulations based on both LTIP3P and LBN3d, with different parameters. 
For the sake of validation, the same minimization problem has also been solved 
by an annealed Metropolis MC method. 
\paragraph{ LTIP3P simulations.--- }
In fig.~\ref{fig:taus}  (upper row) we have shown the total potential energy of the system in  units of KJ/mol vs the LB steps in panel~(a) and their corresponding HBs in  panel~(b) at different $\tau$ but with the same initial random distribution. In the LTIP3P simulations, a hydrogen bond is considered to be formed when the distance between the donor H and the acceptor O is below $2.22$ $\text{A}^\circ$. This value represents  the position of the first minimum of the OH radial  distribution in our simulations,  signaling a bonded state.  
Here, the system size is $N=10^3$ and the damping constant has been set to $\gamma=50 \Delta t^{-1}$. 
Looking at the simulation results, it is clear that in the early time behavior,  only the over-relaxation regime ($\tau/\Delta t=20$) can reach the (global) minimum. The long time behavior of the simulations shows that different $\tau$ finally lead to almost the same minimum, as well as the same HBs,  confirming the consistency of the model. As it can clearly seen from the figure, our results indicate  faster minimization in the over-relaxation regime.  It can also be seen that the HBs increases with the decrease of the potential energy of the system, finally arriving at
$\text{HBs}\approx4$, as expected \cite{kumar2007hydrogen}. 
\paragraph{ LBN3d simulations.---}
A proper choice of parameters aiming  to obtain the desired number of hydrogen bonds ($\text{HBs}\approx4$) was shown to be $\sigma_R=0.28$ and  $\sigma_\theta=0.28$ \cite{Moradi2013}. We should mention that this choice of parameters results in a highly corrugated potential. In fig.~\ref{fig:taus} (lower row), we present the simulation results  (energy in panel~(a)  and HBs in panel~(b)) for a system size of $N=6^3$ with $\gamma=10^5 \Delta t^{-1}$, at different $\tau$, and the same initial random distribution. For the case of LBN3d we report the energy of the system in a convenient dimensionless LB units, i.e., energy per molecule divided by  $\vert\epsilon_{kl}^{HB}\vert$. We note that the optimum value of $\tau$ depends on the potential, which explains why 
the optimum $\tau$ ($\tau/\Delta t=1000$) is different from the case of LTIP3P. 
As in the case of LTIP3P, in the early-time evolution only 
the over-relaxation regime ($\tau/\Delta t=100$ and $\tau/\Delta t=1000$) reaches the minimum, supporting 
faster minimization in the over-relaxation regime. Here, again the HBs increases with decreasing 
potential energy and reaches  up to $\text{HBs}=4$ \cite{Moradi2013}. Under all circumstances, it is seen that the optimum $\tau$ is always much larger than $1$, indicating
that LB operates best far from the hydrodynamic regime.
\begin{figure}
\includegraphics[width=42mm]{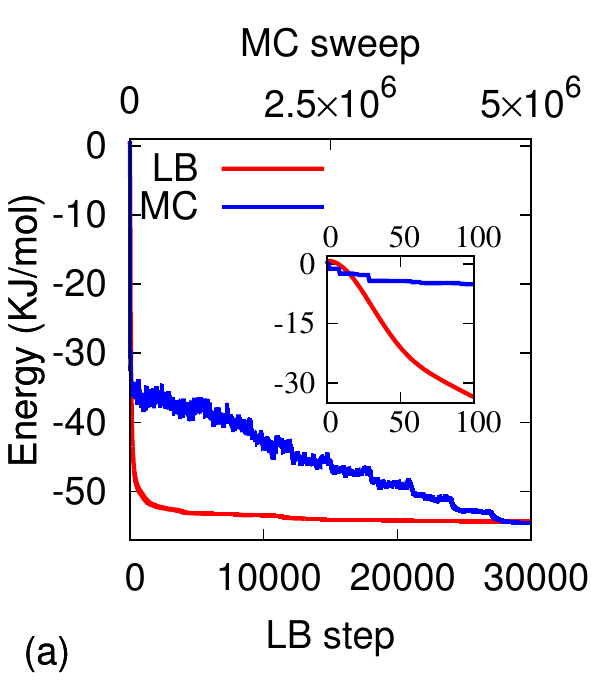}
\includegraphics[width=41mm]{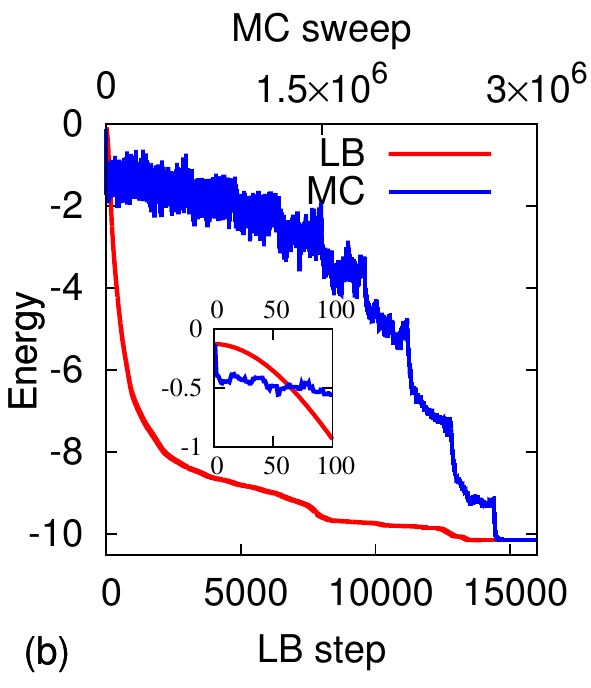}
\caption{Energy vs annealing MC sweeps (in blue) and LB steps  (in red)  for the LTIP3P (a) and the LBN3d (b). 
Both the MC and the LB methods reach the same minimum for the two potentials.  In the case of LTIP3P, the main LB parameters are $N=10^3$, $\gamma=50 \Delta t^{-1}$ and $\tau/\Delta t=20$ and in the case of LBN3d  $N=6^3$, $\gamma=10^5 \Delta t^{-1}$ and $\tau/\Delta t=1000$.
The insets show the very early LB steps/MC sweeps.}
\label{fig:MC} 
\end{figure}
We speculate that non-hydrodynamic operation is more efficient because ``jumps'' are
long enough to avoid trapping in unfavorable  minima, and at the same time, short enough to
avoid overlooking the favorable ones. This is the essence of kinetic over-relaxation: extreme but not too extreme. 
\paragraph{ Comparison with simulated annealing Monte Carlo.---}
In order to validate the model, we have compared our LB results with simulated annealing MC.  In each MC sweep we randomly rotate the molecules, and if the total potential energy of the system $E$ is lower than the previous distribution, the move is accepted.
If not, the move is accepted with probability $p\propto e^{-\beta E}$, where $\beta$ is proportional to the inverse of temperature $T^{-1}$.
In the annealed version,  $\beta$ is progressively increased during the MC simulation.
In the present work, the MC scheme is characterized by three types of moves:
I. global rotations; II. local rotation; III. global refinement.
In I each  molecule rotates about a random axis by a random angle, chosen in the range $[0^\circ,360^\circ]$. 
In  II, we randomly pick one water molecule and rotate it about a random axis by a random angle in the range $[0^\circ,360^\circ]$. III is similar to I, but the random angles are chosen in a narrower range $[0^\circ,10^\circ]$.  In fig.~\ref{fig:MC} we compare  both cases, the LTIP3P and the LBN3d, with the simulated annealing MC discussed above. 
As it can be seen from the figure, in  both cases LB attains the same minimum as in MC, which provides a validation of the LB model.
Although our MC scheme can certainly be improved, the computational performance of the LB minimizer still appears to be pretty remarkable.  
\paragraph{ Time to minimum versus system size.---}
Two important quality factors of dynamic minimizers are 
their robustness towards changes in the initial conditions
and the scaling of the time to solution with the size of the problem.
The LB model is a deterministic minimizer, hence  exposed to a 
sensitivity to the initial conditions for the minimization of highly corrugated potentials. However, this sensitivity can be strongly mitigate by a proper choice of LB parameters.
On the other hand, owing to its strong locality, it features a remarkable linear scaling
with the system size. As a result,  the power of the LB minimizer  is probably best displayed
for large systems, a statement which is only accrued by the excellent amenability
of LB scheme to parallel computing \cite{MUPHY}. As it can be seen from fig.~\ref{fig:MCLB}~(a),  the LB model minimizes 
 water energy based on the LTIP3P  at different system sizes within nearly the same number of LB steps.
Since the computational cost of the LB time-step scales linearly with the system size, we conclude
that CPU time-to-minimum is also linear with the system size.
This stands in notable contrast with annealed MC, where the number of 
sweeps to reach the global minimum grows rapidly with the system size \cite{MCbook2}.   
Since each MC sweep takes about the same CPU time as a LB step, the
end result is that LB becomes increasingly advantageous at larger system sizes. 
It should be further noted that even when LB does not get to the global minimum, due
to a poor choice of the initial condition, it often attains a close-by local minimum.
As a result, one may envisage a synergistic coupling between LB as a fast quasi-global
minimizer, to be combined with a subsequent MC minimization taking the system
to its global minimum. 
\paragraph{Summary.---}
Summarizing, from the sheer view-point of dynamic minimization, the LB
dynamics can be seen as a real-time, extreme over-relaxation dynamics 
in kinetic space.
While standard use of the LB equation is in the hydrodynamic regime, where
it provides a macroscopic solver in``kinetic disguise", for
the purpose of dynamic minimization, it is here found and documented, that
LB provides its best when operated in the non-hydrodynamic regime.  
This unexpected and welcome result might open up a new field of applications
for ``kinetic over-relaxation", as a complement/alternative to Monte Carlo
and Langevin-Molecular-Dynamics to a class of lattice potentials.


\begin{thebibliography}{}


\bibitem{Kirkpatrick1983} S. Kirkpatrick, C. D. Gelatt and M. P. Vecchi,   Science {\bf 220}, 671 (1983).
\bibitem{Marinari1992} E. Marinari and G. Parisi,  Europhys. Lett. {\bf 19}, 451 (1992).
\bibitem{Bryngelson11995} J. D. Bryngelson, J. D. Onuchic, N. D. Socci and P. G.  Wolynes,    Proteins: Structure, Function, and Bioinformatics {\bf 21}, 167 (1995).
\bibitem{Lieb1968} E. H. Lieb and F. Y. Wu,   Phys. Rev. Lett. {\bf 20}, 1445 (1968). 
\bibitem{MCbook1} D. P. Landau and K. Binder,  {\em A Guide To Monte Carlo Simulations In Statistical Physics}, Cambridge University Press, New York (2005).
\bibitem{MCbook2} A. Doucet, N. de Freitas and N. Gordon, {\em Sequential Monte Carlo methods in practice},  Springer,  New York (2001).
\bibitem{Nosea1984} S. Nos\'{e},  Molecular Physics {\bf 52}, 255 (1984).
\bibitem{Car1985} R. Car and M. Parrinello,  Phys. Rev. Lett. {\bf 55}, 2471 (1985).
\bibitem{Rapaport:1985kc} D. C. Rapaport,  Journal of Computational Physics {\bf 60}, 306 (1985).
\bibitem{Jorgensen}  W. L. Jorgensen, J. Chandrasekhar, J. D. Madura, R. W. Impey and M. L. Klein, J. Chem. Phys. {\bf 79}, 926 (1983).
\bibitem{pi2009anomalies} H. L. Pi, J. L. Aragones, C. Vega, E. G. Noya, J.L.F. Abascal, M. A. Gonzalez and C. McBride,   Molecular Physics, {\bf 107}, 365  (2009).
\bibitem{Mark} P. Mark and L. Nilsson,  J. Phys. Chem. A  {\bf 105}, 9954 (2001).
\bibitem{BNbook} A. Ben-Naim,  {\em Molecular Theory of Water and Aqueous Solutions: Part I:  Understanding Water}, World Scientific Publishing Company, Singapore (2010).
\bibitem{Mazzitelli:2011uv} I. Mazzitelli, M. Venturoli, S. Melchionna and S. Succi,  J. Chem. Phys.  {\bf 135}, 124902 (2011).
\bibitem{Moradi2013} N. Moradi, A. Greiner, F. Rao and S. Succi,  J. Chem. Phys.  {\bf 138}, 124105 (2013).
\bibitem{Succi:2001a} S. Succi,  { \em The Lattice Boltzmann Equation}, Oxford University Press, New York (2001).
\bibitem{BSV} R. Benzi, S. Succi, M. Vergassola,  Phys. Rep. {\bf 222}, 145 (1992).
\bibitem{Qian1997} Y.-H. Qian,  Int. J. Mod. Phys. {\bf 8}, 753 (1997).
\bibitem{Zhang2001} R. Zhang, H. Chen, Y. H. Qian  and S. Chen,   PRE {\bf 63}, 056705 (2001).
 \bibitem{Courant}  R. Courant, K. Friedrichs and H. Lewy, IBM Journal of Research and Development, {\bf 11}, 215 (1967).
\bibitem{kumar2007hydrogen} R. Kumar, J. R. Schmidt, and J. L. Skinner,   J. Chem. Phys. {\bf 126}, 204107 (2007).
\bibitem{MUPHY} M. Bernaschi, S. Melchionna, S. Succi, M. Fyta, E Kaxiras, J. K. Sircar,  Computer Physics Communications {\bf 180},  1495 (2009).
\end{thebibliography}
\end{document}